\documentclass[twocolumn,pra,showpacs,superscriptaddress,amssymb,amsmath,amsmath]{revtex4-1}
\usepackage{graphicx}
\usepackage{epstopdf}
\usepackage{bm}
\usepackage{hyperref}
\usepackage{comment}

\hypersetup{%
   pdfpagemode=None, 
   pdfstartpage=1,
   pdfmenubar=true,
   pdftoolbar=true,
   colorlinks = true,
   linkcolor=blue,
   citecolor=blue,
   urlcolor=blue,
   bookmarksopen=false
 }

\newcommand{\be}{\begin{equation}}
\newcommand{\ee}{\end{equation}}

\begin{document}
\title{\textit{Ab initio} properties of the ground-state polar and paramagnetic europium--alkali-metal-atom and europium--alkaline-earth-metal-atom molecules}

\author{Micha\l~Tomza}
\email{michal.tomza@chem.uw.edu.pl}
\affiliation{Faculty of Chemistry, University of Warsaw, Pasteura 1, 02-093 Warsaw, Poland}

\date{\today}

\begin{abstract}

The properties of the electronic ground state of the polar and paramagnetic europium--$S$-state-atom molecules have been investigated. \textit{Ab initio} techniques have been applied to compute the potential energy curves for the europium--alkali-metal-atom, Eu$X$ ($X$=Li, Na, K, Rb, Cs), europium--alkaline-earth-metal-atom, Eu$Y$ ($Y$=Be, Mg, Ca, Sr, Ba), and europium-ytterbium, EuYb, molecules in the Born-Oppenheimer approximation for the high-spin electronic ground state. The spin restricted open-shell coupled cluster method restricted to single, double, and noniterative triple excitations, RCCSD(T), was employed and the scalar relativistic effects within the small-core energy-consistent pseudopotentials were included. The permanent electric dipole moments and static electric dipole polarizabilities were computed. The leading long-range coefficients describing the dispersion interaction between atoms at large internuclear distances $C_6$ are also reported.
The EuK, EuRb, and EuCs molecules are examples of species possessing both large electric and magnetic dipole moments making them potentially interesting candidates for ultracold many-body quantum simulations when confined in an optical lattice in combined electric and magnetic fields.

\end{abstract}

\pacs{34.20.-b, 33.15.Kr, 31.50.Bc}

\maketitle

\section{Introduction}

The creation of the ultracold dense gas of polar molecules is an important research goal since the heteronuclear molecules possessing a permanent electric dipole moment are promising candidates for numerous applications ranging from the ultracold controlled chemistry to the quantum simulations of many-body physics~\cite{ColdMolecules,CarrNJP09,QuemenerCR12}.
Polar alkali-metal dimers have already been produced in their absolute rovibrational ground state close to the quantum degeneracy~\cite{NiScience08} and employed in many groundbreaking experiments~\cite{OspelkausScience10,NiNature10,BoNature13}. Motivated by the promise of exciting applications, the creation of ultracold polar molecules with a spin structure and a magnetic dipole moment is currently emerging as another important research goal~\cite{MicheliNatPhys06,RablPRA07,KremsNJP10,BaronScience2014}.

The simplest diatomic molecules with a spin structure and a magnetic dipole moment are those in the electronic state of the $^2\Sigma$ and $^3\Sigma$ symmetry. The former ones can consist of a closed-shell atom such as Sr or Yb and an open-shell atom such as Rb or F, whereas the latter ones can be found in the alkali-metal dimers. Until now, ultracold dense gas of the $^2\Sigma$ state molecules in the rovibrational ground state has not yet been produced 
albeit the efforts being undertaken to direct laser cool molecules such as SrF~\cite{ShumanNature10} or YO~\cite{HummonPRL13} and photo- or magnetoassociate molecules such as RbYb~\cite{NemitzPRA09} and RbSr~\cite{SchreckPRA13}. The magnetoassociation of the $^2\Sigma$ state molecules, though possible, is rendered very difficult due to the small resonance widths~\cite{ZuchowskiPRL10,BruePRL12,TomzaPRL14}. The ultracold dense gases of alkali metal-dimers in the $^3\Sigma$ state have already been produced~\cite{NiScience08,DenschlagPRL08} but they may undergo chemical reactions and the use of an optical lattice to segregate the molecules and suppress losses may be necessary~\cite{TomzaPRA13b}.

Recently, more and more atoms with a complex electronic structure have been cooled down and investigated at low or ultralow temperatures. On one hand, the Fesbach resonances in the mixtures of the Yb atoms in the metastabe $^3P_2$ state with the ground state Yb~\cite{KatoPRL13} or Li~\cite{KhramovPRL14} atoms have been explored. On the other hand, the Bose-Einstein condensate or Fermi degeneracy of highly magnetic atoms such as Cr~($^7S$)~\cite{PfauPRL05}, Er~($^3H$)~\cite{FerlainoPRL12,AikawaPRL14}, or Dy~($^5I$)~\cite{LevPRL11,LevPRL12} have been obtained and the Fano-Feshbach resonances were observed in the ultracold gas of the Er atoms~\cite{FrischNature14}. This opens the way for the formation of ultracold molecules containing these atoms.

The chromium--alkali-metal-atom molecules such as CrRb~\cite{SadeghpourPRA10} and the chromium--closed-shell-atom molecules such as CrSr or CrYb~\cite{TomzaPRA13a} have been proposed as candidates for molecules with both large magnetic and electric dipole moments.
The molecules formed from the Er or Dy atom and the alkali-metal or alkaline-earth-metal atom may be polar and will have a large magnetic dipole moment. 
Unfortunately, the density of the interaction-anisotropy-induced Feshbach resonances between the lanthanide atoms with the large electronic orbital angular momentum is very high~\cite{KotochigovaPRL12} and the chaotic behavior in the interaction between these atoms~\cite{FrischNature14} may make the magnetoassociation of heteronuclear molecules difficult.
At the same time, the density of the Feshbach resonances between the highly magnetic Eu atom in the spherically symmetric $^8S$ electronic ground state will be lower~\cite{SuleimanovPRA10}, and molecules containing this atom may also be polar and will have a large magnetic dipole moment. 

The buffer-gas cooling and magnetic trapping of the Eu atoms were demonstrated~\cite{DoylePRL97,DoyleNature04} and the further cooling to the quantum degeneracy will not be more challenging than the already demonstrated production of the ultracold gas of the similar Cr atoms~\cite{PfauPRL05} or other lanthanide atoms with a more complex electronic structure~\cite{LevPRL11,LevPRL12,FerlainoPRL12,AikawaPRL14}. 

The interaction of the ${}^8S$ state Eu atom with the ${}^2S$ state alkali-metal atom gives rise to the two molecular electronic states of the $X^7\Sigma^-$ and $a^9\Sigma^-$ symmetries. 
The interaction of the ${}^8S$ state Eu atom with the  ${}^1S$ closed-shell alkaline-earth-metal atom results in one electronic state of the $X^8\Sigma^-$ symmetry only~\cite{Herzberg}.
The molecules in the $X^8\Sigma^-$ electronic state inherit the large magnetic dipole moment of the Eu atom $d_m=7\,\mu_B$ ($\mu_B$ is the Bohr magneton), whereas the magnetic dipole moment of the molecules in the $X^7\Sigma^-$ and $a^9\Sigma^-$ electronic states is $6\,\mu_B$ and $8\,\mu_B$, respectively.  

In the present work we investigate the properties of the high-spin electronic ground state of the
europium--alkali-metal-atom, europium--alkaline-earth-metal-atom, and europium-ytterbium molecules. 
To the best of our knowledge, these molecules have not yet been considered theoretically or experimentally.
Here we fill this gap and report the \textit{ab initio} properties of the $a^9\Sigma^-$ electronic state of the europium--alkali-metal-atom molecules and the $X^8\Sigma^-$ electronic ground state of the europium--alkaline-earth-metal-atom and europium-ytterbium molecules paving the way towards a more elaborate study of the formation and application of these polar and paramagnetic molecules.

The plan of our paper is as follows. Section~\ref{sec:theory} describes the theoretical methods used in the \textit{ab initio} calculations. Section~\ref{sec:results} discusses the potential energy curves and electric  properties of the europium--$S$-state-atom molecules in the rovibrational ground state.
Section~\ref{sec:summary} summarizes our paper.

\section{Computational details}
\label{sec:theory}

The electronic configuration of the Eu atom in the electronic ground state is $[\mathrm{Xe}]~4f^76s^2$ and the corresponding term is $^8S$. The $4f$ shell is symmetrically  half filled and responsible for the open-shell character of the Eu atom whereas the outermost $6s$ shell is closed. 
Buchachenko \textit{et al.}~\cite{BuchachenkoJCP09} have demonstrated that
 the interaction between the $4f$ electrons of the Eu atoms in the Eu$_2$ dimer is extremely weak. 
This suggests that the $4f$ electrons are screened and the interaction between the Eu and other atom will be dominated by the outermost valence electrons.
Nevertheless, the $4f$ electrons have to be carefully included in any accurate computational model. This  makes any \textit{ab initio} calculations involving lanthanide atoms very challenging~\cite{BuchachenkoEPJD07,ZhangJCP10,GopakumarJCP10,TomzaPCCP11,KotochigovaPRL12,BorkowskiPRA13}. 
  
We have calculated the potential energy curves for the high-spin electronic ground state of the europium--$S$-state-atom molecules in the Born-Oppenheimer approximation by using the spin-restricted open-shell coupled cluster method restricted to single, double, and noniterative triple excitations, starting from the restricted open-shell Hartree-Fock (ROHF) orbitals, RCCSD(T)~\cite{KnowlesJCP99}. The interaction energies were obtained by the supermolecule method and the Boys and Bernardi scheme was utilized to correct for the basis-set superposition error~\cite{BoysMP70}.

The scalar relativistic effects were accounted for by using small-core relativistic energy-consistent pseudopotentials (ECP) to replace the inner-shells electrons~\cite{DolgCR12}. The use of the pseudopotentials 
 allowed to use larger basis sets to describe the valence electrons and modeled the inner-shells electrons density as accurately as the high quality atomic calculation used to fit the pseudopotentials.
Note that 35 electrons ($4s^24p^64d^{10}4f^75s^25p^66s^2$) in the Eu atom were correlated in all calculations. 

The Li, Na, Be, and Mg atoms were described by the augmented correlation consistent polarized valence quadruple-$\zeta$ quality basis sets (aug-cc-pVQZ)~\cite{DunningJCP89}.
The K, Rb, Cs, Ca, Ba, Yb, and Eu atom were described by the pseudopotentials from the Stuttgart library. 
The K, Rb, and Cs atoms were described with the ECP10MDF, ECP28MDF, and ECP46MDF pseudopotentials~\cite{LimJCP06}, and $[11s11p5d3f]$, $[14s14p7d6f1g]$, and $[14s11p6d4f2g]$ basis sets obtained by augmenting the basis sets suggested by the authors of Ref.~\cite{LimJCP06}, respectively.
The Ca, Sr, and Ba atoms were described with the ECP10MDF, ECP28MDF, and ECP46MDF pseudopotentials~\cite{LimJCP06} and $[12s12p7d3]$, $[14s11p6d5f4g]$, and $[13s12p6d5f4g]$ basis sets obtained by augmenting the basis sets suggested by the authors of Ref.~\cite{LimJCP06}, respectively.
The Eu and Yb atom were described with the ECP28MDF pseudopotentials~\cite{DolgTCA98} and $[15s15p11d10f6g]$ and $[15s14p12d11f8g]$ basis sets~\cite{DolgTCA98} augmented with additional diffuse functions~\cite{BuchachenkoSC07}. 
In all calculations the basis sets were augmented by the set of $[3s3p2d1f1g]$ bond functions~\cite{midbond}.

The permanent electric dipole moments
and static electric dipole polarizabilities
were calculated by using the finite field method. 
The $z$ axis was chosen along the internuclear axis and oriented from the alkali-metal or alkaline-earth-metal atom to the
Eu atom.

The interaction potential between two neutral atoms in the electronic ground state is asymptotically given by the dispersion interaction of the form  $-{C_6}/{R^6}$
where the $C_6$ coefficient can be calculated as the integral over the product of the dynamic polarizabilities of two atoms at an imaginary frequency, $C_6=\frac{\pi}{3}\int_0^\infty \alpha_{\textrm{A}}(i\omega)\alpha_\textrm{B}(i\omega)d\omega$~\cite{HeijmenMP96}. 

The dynamic electric dipole polarizabilities at an imaginary frequency $\alpha(i\omega)$ of the  alkali-metal and alkaline-earth-metal atoms were taken from the work by Derevianko \textit{et al.}~\cite{DerevienkoADNDT10}, whereas the dynamic polarizabilities of the Eu and Yb atoms were obtained by using the explicitly connected representation of the expectation value and polarization propagator within the coupled cluster method~\cite{MoszynskiCCCC05} and the best approximation XCCSD4 proposed by Korona and collaborators~\cite{KoronaMP06}. The Eu atom in the calculation of the dynamic polarizabilities was described by the large-core relativistic energy-consistent pseudopotential ECP53MWB~\cite{DolgTCA93} combined with the core-polarization potential~\cite{DolgTCA98} and additionally augmented basis set~\cite{HulsenTCA11,BuchachenkoSC07}.

The multireference configuration interaction method restricted to single and double excitations (MRCISD)  was employed to calculate the energy splitting between the potential energy curves of the $X^7\Sigma^-$ and $a^9\Sigma^-$ electronic states of the europium--alkali-metal-atom molecules.

In the present work we have adopted the computational scheme successfully applied to the ground and excited states of the SrYb molecule~\cite{TomzaPCCP11}, alkali-metal dimers in the $a^3\Sigma^+_u$ state~\cite{TomzaPRA13b}, and chromium--closed-shell-atom molecules~\cite{TomzaPRA13a}.
The employed CCSD(T) method and basis sets reproduce the potential well depths of the alkali-metal and alkaline-earth-metal dimers with an error of a few percent ($\lesssim5\%$) as compared to the experimental results~\cite{TomzaMP13,SkomorowskiJCP12} and an error of the atomic polarizabilities is even smaller.
Nevertheless, the calculations for the molecules containing an open-shell lanthanide atom are more challenging. 
Based on the present convergence analysis, we estimate that the total uncertainty of the calculated potential energy curves and electronic properties is of the order of 20\% and the present model sets a lower bound for the binding energy.
The lack of the exact treatment of the triple and higher excitations in the employed CCSD(T) method is a preliminary limiting factor~\cite{TomzaPRA13a}.

\begin{figure}[t!]
\begin{center}
\includegraphics[width=0.95\columnwidth]{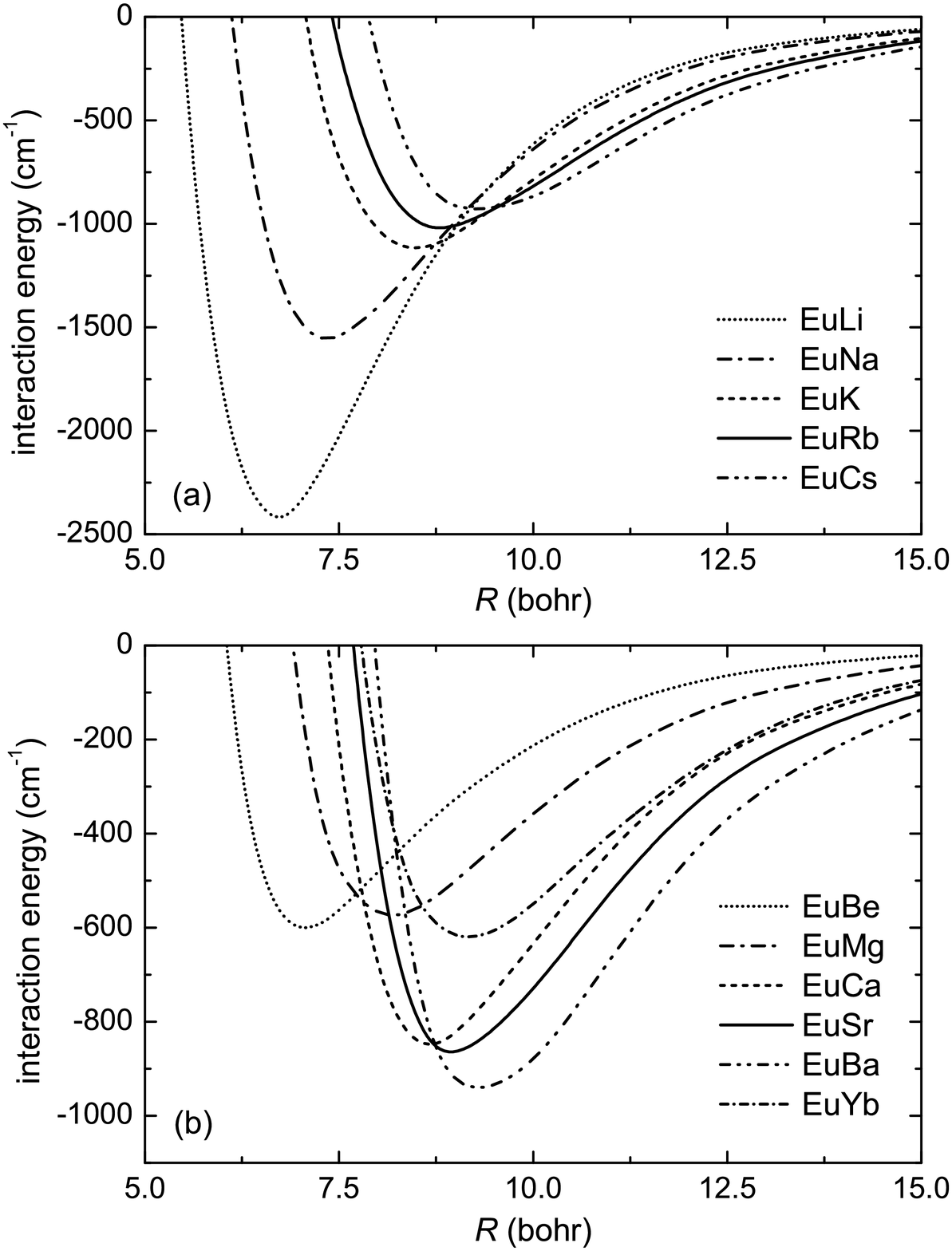}
\end{center}
\caption{Potential energy curves of the $a^9\Sigma^-$ electronic state of the EuLi, EuNa, EuK, EuRb, and EuCs molecules (a) and of the $X^8\Sigma^-$ electronic ground state of the EuBe, EuMg, EuCa, EuSr, EuBa, and EuYb molecules (b).}
\label{fig:Curves}
\end{figure}

All calculations were performed with the \textsc{Molpro} package of \textit{ab initio} programs \cite{Molpro}

\section{Numerical results and discussion}
\label{sec:results}

\begin{table*}[t!]
\caption{Spectroscopic characteristics: Equilibrium bond length $R_e$, well depth $D_e$, harmonic frequency $\omega_0$, number of bound vibrational states $N_v$, and long-range dispersion coefficient $C_6$ of the high-spin electronic ground state and rotational constant $B_0$, electric dipole moment $d_0$, average polarizability $\bar{\alpha}_0$, and polarizability anisotropy $\Delta\alpha_0$, for the rovibrational ground level 
of the $a^9\Sigma^-$ electronic state of the EuLi, EuNa, EuK, EuRb, and EuCs molecules and of the $X^8\Sigma^-$ electronic ground state of the EuBe, EuMg, EuCa, EuSr, EuBa, and EuYb molecules. 
For the EuLi, EuNa, EuK, EuRb, and EuCs molecules the difference of the well depth of the $a^9\Sigma^-$ and $X^7\Sigma^-$ electronic states $\Delta D_e$.
Masses of the most abundant isotopes are assumed. }
\begin{ruledtabular}
\begin{tabular}{lrrrrrrrrrr}
Molecule & $R_e$ (bohr) & $D_e$ (cm$^{-1}$) &  $\omega_0$(cm$^{-1}$) & $N_v$ & $B_0$(cm$^{-1}$)&  $d_0$(D) & $\bar\alpha_0$(a.u.) & $\Delta\alpha_0$(a.u.) & $C_6$(a.u.) & $\Delta D_e$ (cm$^{-1}$)  \\
\hline 
EuLi & 6.73 & 2443 & 223 & 21 & 0.198 & -0.16 & 394 & 364 & 2066 & 605 \\ 
EuNa & 7.36 & 1567 & 79  & 46 & 0.056 & -0.54 & 530 & 360 & 2220 & 280 \\ 
EuK  & 8.41 & 1141 & 49  & 54 & 0.027 & -1.23 & 547 & 519 & 3434 & 229 \\ 
EuRb & 8.79 & 1047 & 35  & 71 & 0.014 & -1.41 & 580 & 539 & 3779 & 199 \\ 
EuCs & 9.28 &  983 & 28  & 80 & 0.010 & -1.67 & 670 & 634 & 4535 & 161\\ 
EuBe & 7.04 &  605 & 87  & 18 & 0.143 &  0.45 & 259 & 163 &  817 & - \\
EuMg & 8.11 &  587 & 45  & 31 & 0.044 &  0.10 & 295 & 190 & 1423 & - \\
EuCa & 8.62 &  870 & 44  & 47 & 0.026 &  0.17 & 411 & 343 & 2648 & - \\
EuSr & 8.95 &  897 & 33  & 65 & 0.013 &  0.12 & 452 & 373 & 3202 & - \\
EuBa & 9.29 & 1000 & 28  & 79 & 0.010 &  0.08 & 548 & 464 & 4112 & - \\
EuYb & 9.13 &  632 & 22  & 68 & 0.009 & -0.13 & 364 & 242 & 2727 & - \\
\end{tabular}
\label{tab:Spectro}
\end{ruledtabular}
\end{table*}

\subsection{Potential energy curves}

The computed potential energy curves of the $a^9\Sigma^-$ electronic state of the EuLi, EuNa, EuK, EuRb, and EuCs molecules are presented in Fig.~\ref{fig:Curves}(a) and the potential energy curves of the $X^8\Sigma^-$ electronic ground state of the EuBe, EuMg, EuCa, EuSr, EuBa, and EuYb molecules are presented in Fig.~\ref{fig:Curves}(b). The corresponding long-range $C_6$ coefficients are reported in Table~\ref{tab:Spectro}.
The equilibrium distances $R_e$ and well depths $D_e$ are also collected in Table~\ref{tab:Spectro}.

\begin{figure}[t!]
\begin{center}
\includegraphics[width=0.95\columnwidth]{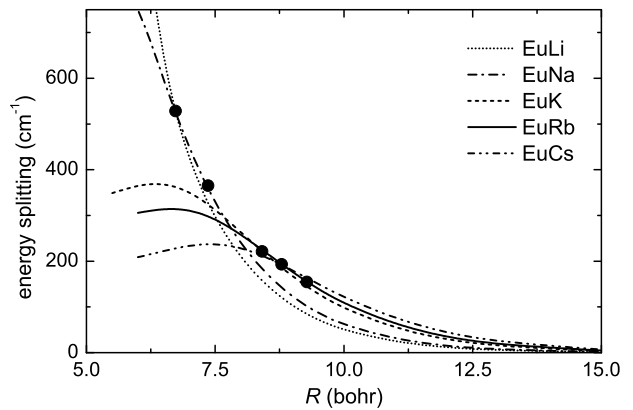}
\end{center}
\caption{Energy splittings between the $X^7\Sigma^-$ and $a^9\Sigma^-$ electronic states of the EuLi, EuNa, EuK, EuRb, and EuCs molecules. Points indicate the values for the equilibrium distance of the  $a^9\Sigma^-$ state.}
\label{fig:split}
\end{figure}

An inspection of Fig.~\ref{fig:Curves} reveals that all potential energy curves show a smooth behavior with well defined minima. The well depths for the europium--alkali-metal-atom molecules are slightly larger than for the europium--alkaline-earth-metal-atom molecules. For the former ones the largest well depth is 2443$\,$cm$^{-1}$ for the EuLi molecule and the smallest one is 983$\,$cm$^{-1}$ for the EuCs molecule. 
For the latter ones  the largest well depth is 1000$\,$cm$^{-1}$ for the EuBa molecule and the smallest one is 587$\,$cm$^{-1}$ for the EuMg molecule. 
Interestingly, the well depths of the europium--alkali-metal-atom molecules are systematically
decreasing with the increasing mass of the alkali-metal atom whereas the well depths of the europium--alkaline-earth-metal-atom molecules are systematically
increasing with the increasing mass of the alkaline-earth-metal atom, except for the EuBe molecule (in accordance with the results for the CrBe~\cite{TomzaPRA13a} and Be$_2$~\cite{PatkowskiScience09} molecules). The different trends can be explained by analyzing the bonding of the atoms as a result of the interaction of the electrons from the outermost $s$ shells only in the molecular orbitals picture. The formal order of the chemical bond is equal to half for the europium--alkali-metal-atom molecules and zero for the europium--alkaline-earth-metal-atom molecules. For this reason the former ones are chemically bound whereas the latter ones are stabilized by the dispersion interaction only.

\begin{figure}[t!]
\begin{center}
\includegraphics[width=0.95\columnwidth]{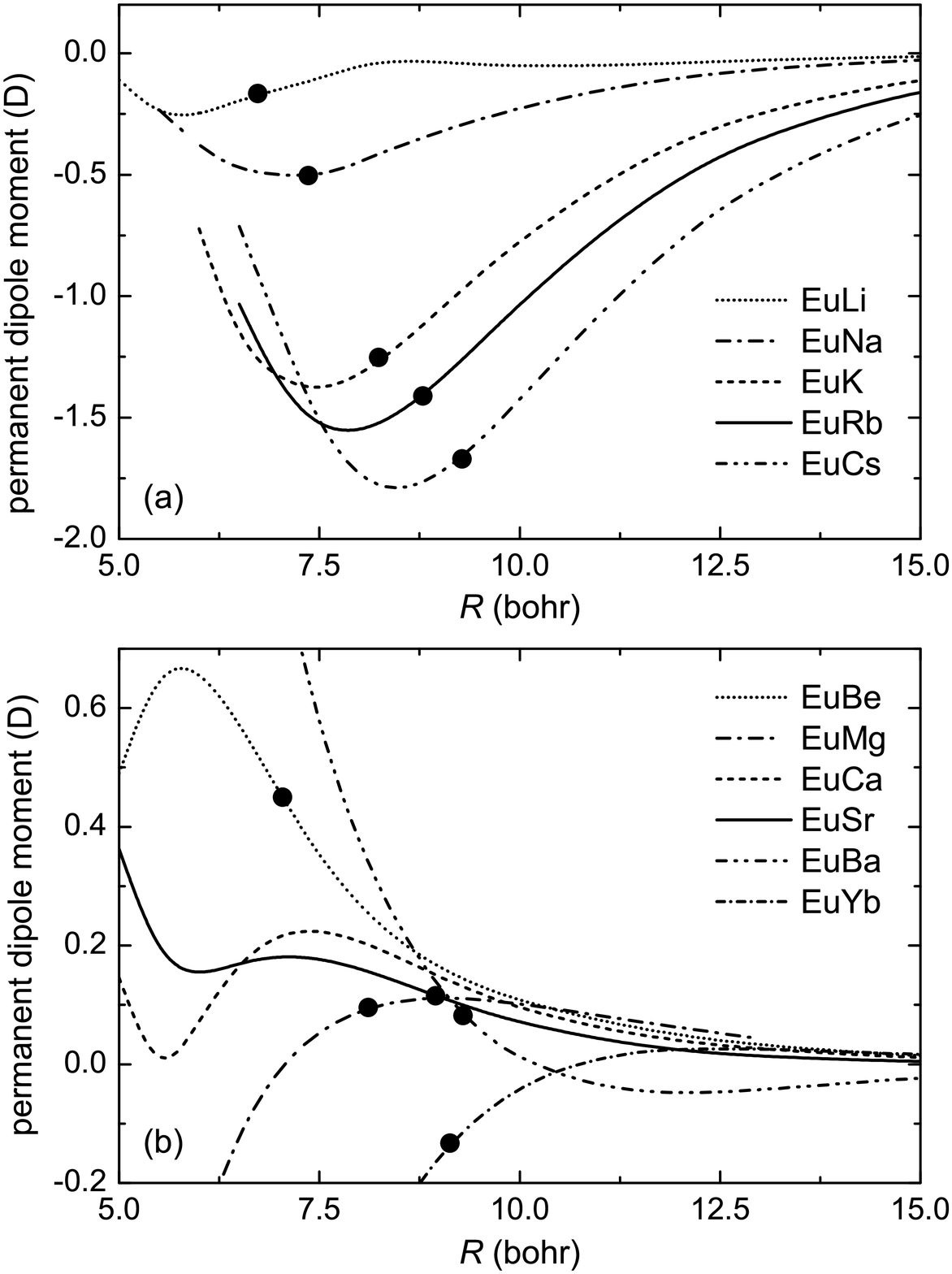}
\end{center}
\caption{Permanent electric dipole moments of the $a^9\Sigma^-$ electronic state of the EuLi, EuNa, EuK, EuRb, and EuCs molecules (a) and of the $X^8\Sigma^-$ electronic ground state of the EuBe, EuMg, EuCa, EuSr, EuBa, and EuYb molecules (b). Points indicate the values for the ground rovibrational level.}
\label{fig:Dip}
\end{figure}

The equilibrium distances for the europium--alkali-metal-atom molecules take values between 6.73 bohr for the EuLi molecule up to 9.28 bohr for the EuCs molecule and for the europium--alkaline-earth-metal-atom molecules take values between 7.04 bohr for the EuBe molecule up to 9.29 bohr for the EuBa molecule. 
The equilibrium distances of all investigated molecules are systematically increasing with the increasing mass of the alkali-metal or alkaline-earth-metal atom. 

\textit{Ab initio} potentials were used to calculate the rovibrational spectra of the $a^9\Sigma^-$ and $X^8\Sigma^-$ electronic states for the molecules consisting of the most abundant isotopes. The harmonic frequencies $\omega_0$ and the numbers of the supported bound states for the angular momentum $J=0$, $N_\upsilon$,  are reported in Table~\ref{tab:Spectro}. The rotational constants $B_0$ for the rovibrational ground state $v=0,J=0$ were also calculated and are reported in Table~\ref{tab:Spectro}.

The energy splittings between the $X^7\Sigma^-$ and $a^9\Sigma^-$ electronic states of the EuLi, EuNa, EuK, EuRb, and EuCs molecules as functions of the internuclear distance $R$ are presented in Fig.~\ref{fig:split}. They show a smooth behavior, take a value of a few hundred cm$^{-1}$ around the equilibrium bond length, and for large internuclear distances tend exponentially to zero.
They were calculated with the MRCISD method which is less accurate than the RCCSD(T) method. Nevertheless, the  energy splitting between the considered states is dominated by the exchange energy~\cite{HeitlerZP27}, which, for present purposes, is reproduced enough accurately already at the mean field level of the theory. Therefore, the potential energy curves of the $X^7\Sigma^-$ electronic  ground state of the europium--alkali-metal-atom molecules, as accurate as the present ones of the $a^9\Sigma^-$ state, can be obtained by subtracting the energy splittings from the potential energy curves of the $a^9\Sigma^-$ electronic state calculated with the RCCSD(T) method.   
The differences between the well depths of the $X^7\Sigma^-$ and $a^9\Sigma^-$  electronic states are reported in Table~\ref{tab:Spectro}. The well depths of the $X^7\Sigma^-$ state are larger by values between $14\,\%$ for the EuCs molecule up to $20\,\%$ for the EuLi molecule.
The equilibrium bond lengths of this state are smaller by 0.15-0.5$\,$bohr.
The above results suggest that the interaction between the electrons of the open $f$-shell of the lanthanide atom with the electron of the open $s$-shell of the alkali-metal atom can not be neglected and the $f$-shell electrons of the open-shell lanthanide atom have to be implicitly included in any accurate computational scheme.

\begin{figure}[t!]
\begin{center}
\includegraphics[width=0.95\columnwidth]{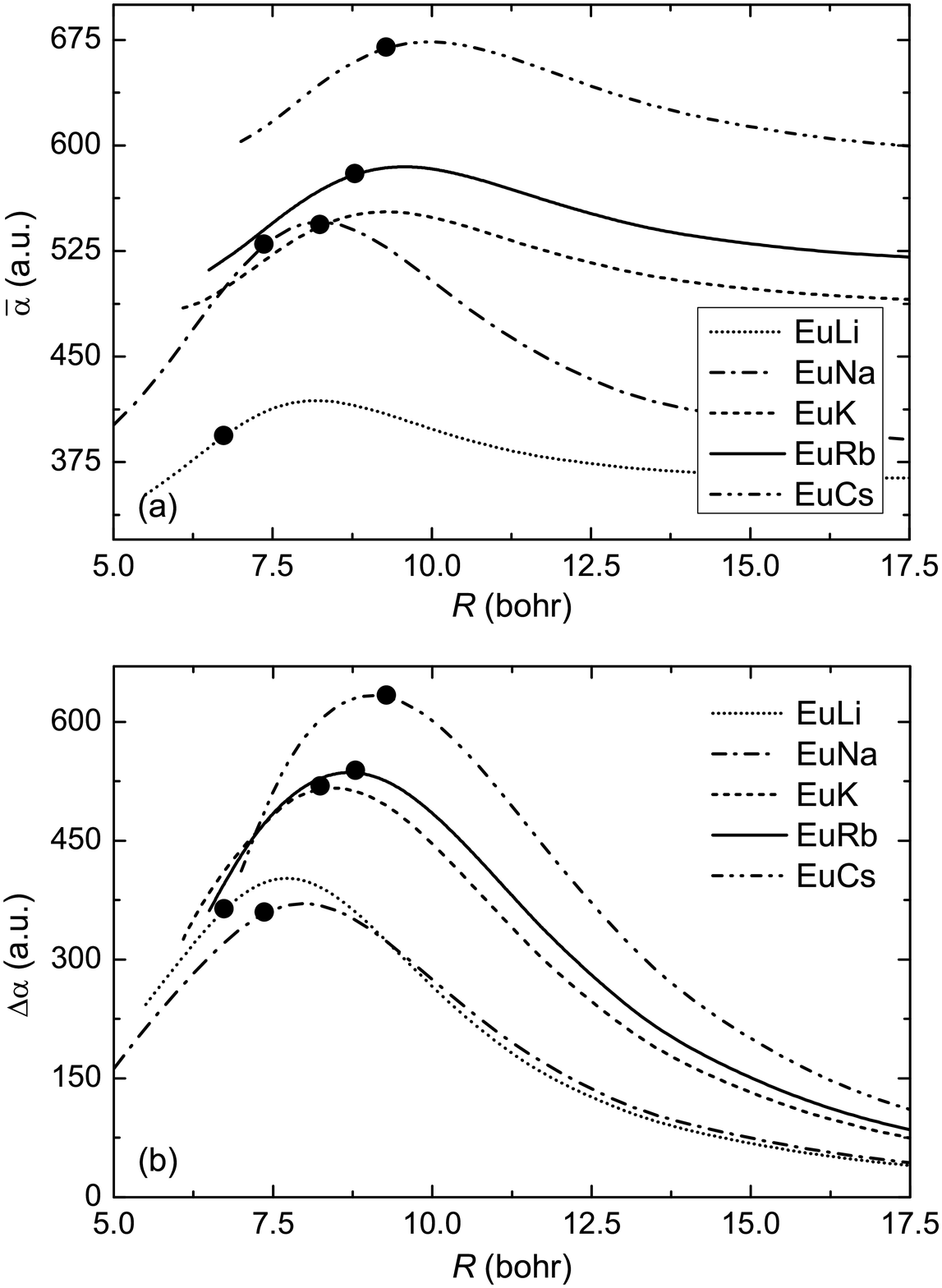}
\end{center}
\caption{The average polarizability (a) and polarizability anisotropy (b) of the $a^9\Sigma^-$ electronic state of the EuLi, EuNa, EuK, EuRb, and EuCs molecules. Points indicate the values for the ground rovibrational level.}
\label{fig:polariAM}
\end{figure}
\begin{figure}[t!]
\begin{center}
\includegraphics[width=0.95\columnwidth]{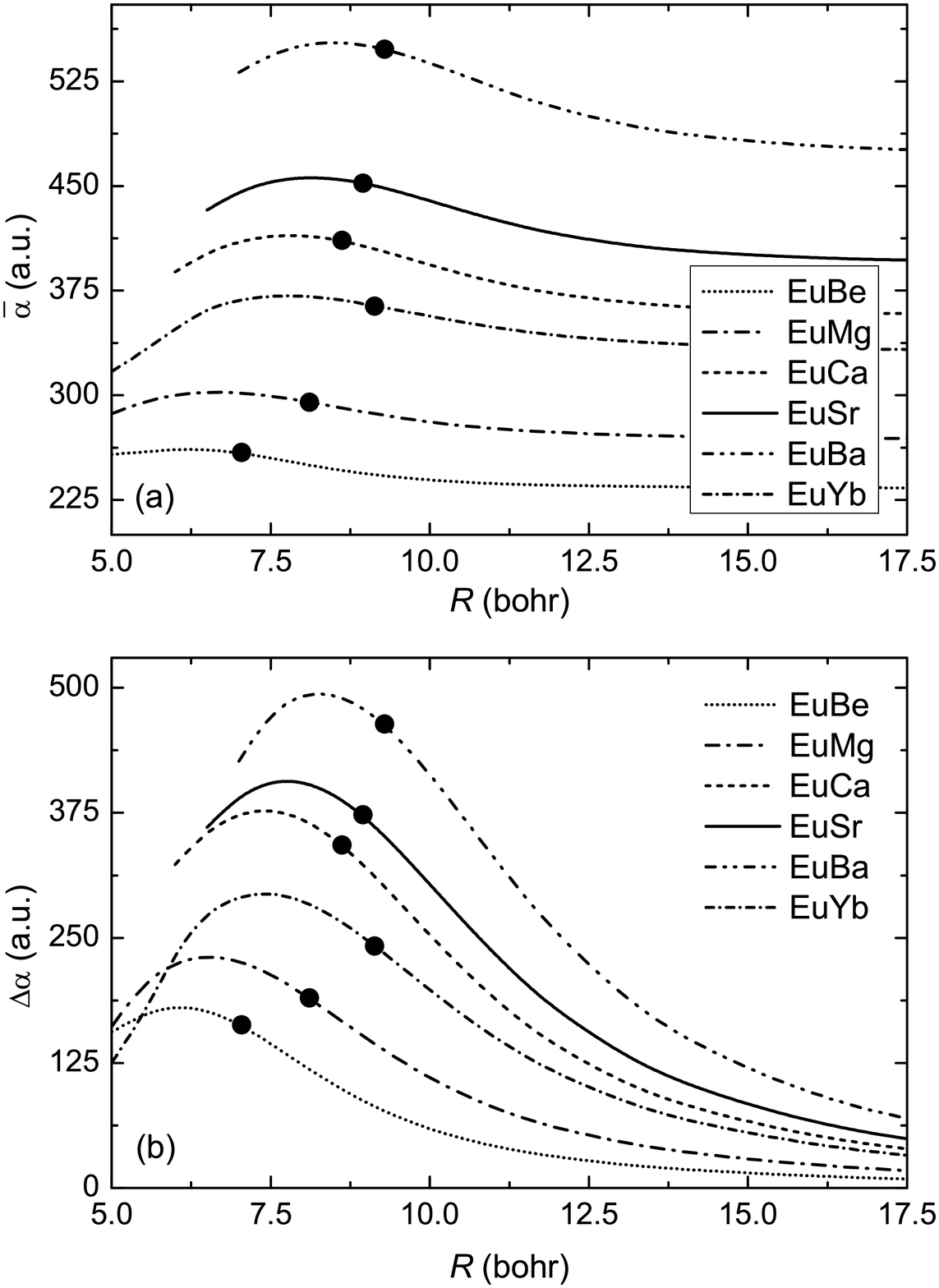}
\end{center}
\caption{The average polarizability (a) and polarizability anisotropy (b) of the $X^8\Sigma^-$ electronic ground state of the  EuBe, EuMg, EuCa, EuSr, EuBa, and EuYb molecules. Points indicate the values for the ground rovibrational level.}
\label{fig:polariAEM}
\end{figure}

\subsection{Permanent electric dipole moments and static electric dipole polarizabilities}

The many-body dynamics of the molecules at ultralow temperatures can be manipulated and controlled with the external static electric or nonresonant laser fields which couple with the permanent electric dipole moment and  electric dipole polarizability, respectively~\cite{QuemenerCR12,LemeshkoMP13}. The static electric field orients the molecules whereas the nonresonant laser field align them~\cite{StapelfeldtRMP03}. Both fields can be used to control the rovibrational structure~\cite{TomzaMP13}, the magnetic Feshbach resonances and magnetoassociation~\cite{KremsPRL06,TomzaPRL14}, as well as the shape resonances and photoassociation~\cite{GonzalezPRA12}.
All applications of molecules confined in optical lattices rely on the long-range dipole-dipole interactions therefore the magnitude of the permanent electric dipole moment is a figure of merit for experiments with lattice-confined molecules~\cite{ColdMolecules,CarrNJP09,QuemenerCR12}.

The permanent electric dipole moments of the EuLi, EuNa, EuK, EuRb, and EuCs molecules in the $a^9\Sigma^-$ electronic state as functions of the internuclear distance $R$ are presented in Fig.~\ref{fig:Dip}(a) and the permanent electric dipole moments of the EuBe, EuMg, EuCa, EuSr, EuBa, and EuYb molecules in the $X^8\Sigma^-$ electronic ground state are presented in Fig.~\ref{fig:Dip}(b). 
The values for the ground rovibrational level are reported in Table~\ref{tab:Spectro}.

The EuK, EuRb, and EuCs molecules have the largest permanent electric dipole moments. Their values for the rovibrational ground state are 1.23~D, 1.41~D, and 1.67~D, respectively. These values are two times larger than 0.6~D of the KRb molecule~\cite{AymarJCP05}, and similar to 1.2~D of the RbCs molecule~\cite{AymarJCP05}, 1.36~D of the RbSr molecule~\cite{ZuchowskiPRL10}, 1.48~D of the CrSr molecule~\cite{TomzaPRA13a}, and 1.19~D of the CrYb molecule~\cite{TomzaPRA13a}.
Other investigated europium--$S$-state-atom molecules have a permanent electric dipole moment much smaller (below 0.5~D) and thus are less promising candidates for the realization of ultracold dipolar gases.

The average polarizability $\bar{\alpha}=(2\alpha_{\perp}+\alpha_{\parallel})/3$, where
$\alpha_{\perp}$ and $\alpha_{\parallel}$ are the perpendicular and parallel components of the polarizability tensor, and the polarizability anisotropy $\Delta\alpha=\alpha_{\parallel}-\alpha_{\perp}$ of the EuLi, EuNa, EuK, EuRb, and EuCs molecules in the $a^9\Sigma^-$ electronic state are presented in Fig.~\ref{fig:polariAM} and the average polarizability $\bar{\alpha}$ and the polarizability anisotropy $\Delta\alpha$ of the EuBe, EuMg, EuCa, EuSr, EuBa, and EuYb molecules in the $X^8\Sigma^-$ electronic ground state are presented in Fig.~\ref{fig:polariAEM}. The values for the ground rovibrational level are reported in Table~\ref{tab:Spectro}.

The polarizabilities show an overall smooth behavior and tend smoothly to their asymptotic atomic values. The interaction-induced variation of the polarizability is clearly visible while changing the internuclear distance $R$. The polarizability anisotropies for the rovibrational ground state of the investigated molecules are larger than the ones of the alkali-metal-atom dimers~\cite{DeiglmayrJCP08} and chromium--closed-shell-atom molecules~\cite{TomzaPRA13a}. Therefore the control and alignment with the nonresonant laser field will be facilitated~\cite{GonzalezPRA12,TomzaMP13,TomzaPRL14}.
In the present work, we have calculated static polarizabilities which describe the interaction of molecules with the far nonresonant field. When the shorter-wavelength field is applied the dynamic
polarizabilities have to be used, which usually are larger but of the same order of magnitude as the static ones.


\section{Summary and conclusions}
\label{sec:summary}

In the present work we have investigated the \textit{ab initio} properties of the europium--alkali-metal-atom, europium--alkaline-earth-metal-atom, and europium--ytterbium molecules. Potential energy curves, permanent electric dipole moments, and static electric dipole polarizabilities for the molecules in the electronic ground state were obtained with the spin-restricted open-shell coupled cluster method restricted to single, double, and noniterative triple excitations, RCCSD(T), in the Born-Oppenheimer approximation. The scalar relativistic effects within energy-consistent pseudopotentials were included. The properties of the molecules in the rovibrational ground state were analyzed. The leading long-range coefficients describing the dispersion interaction between the atoms at large internuclear distances, $C_6$, were also computed.

The molecules under investigation are examples of species possessing both large magnetic and electric dipole moments. Especially the EuK, EuRb, and EuCs molecules in the rovibrational ground state have the permanent electric dipole moment as large as 1.23~D, 1.41~D, and 1.67~D, respectively.
This makes them potentially interesting candidates for ultracold collisional studies of
dipolar molecules in the combined electric and magnetic fields when the magnetic dipole-dipole interaction can compete with the electric dipole-dipole interaction. The combination of the large electric and magnetic dipole moments with the spin structure opens the way for new potentially interesting applications of these molecules when confined in an optical lattice at ultralow temperatures~\cite{MicheliNatPhys06}. In addition to the rotational structure and electric dipole moment, the large electronic spin $S=3$ or $4$ can be employed to realize a multi-qubit spin register, on one hand, and the related magnetic dipole moment (6-8$\,\mu_B$) will rise for the direct nearest-neighbor interaction of the order of 5-50$\,$Hz for typical optical lattices, on the other hand.
However, the actual many-body properties of the ultracold highly magnetic and polar molecules trapped in an optical lattice in the combined electric and magnetic fields still have to be investigated. 

The investigated molecules also inherit an interesting and rich hyperfine structure of the Eu atom resulting from the coupling of the nuclear spin $I=5/2$ with the electronic spin $S=7/2$~\cite{SandarsPRSA60}. In the europium--alkaline-earth-metal-atom and europium--ytterbium molecules the splitting of the six Eu hyperfine levels ($F=1-6$) will be slightly modified by the presence of the bosonic closed-shell atom~\cite{ZuchowskiPRL10} or new levels will be induced 
 by the interaction with the nuclear spin of the fermionic closed-shell atom~\cite{BruePRL12}.
In the case of the  europium--alkali-metal-atom molecules the interaction between the total electronic spin with the nuclear spins of the Eu and alkali-metal atoms will result in a new molecular hyperfine structure.  In a magnetic field, all hyperfine levels will split into many Zeeman sublevels.

The formation of the investigated molecules is out of the scope of the present work. Nevertheless, the europium--alkali-metal-atom molecules can be formed in the same manner as the alkali-metal dimers i.e.~by using the magnetoassociation within the vicinity of the Feshbach resonances followed by the stimulated Raman adiabatic passage (STIRAP)~\cite{NiScience08}. The europium--alkaline-earth-metal-atom and europium--ytterbium molecules can potentially be magnetoassociated by employing the Feshbach resonances caused by the interaction-induced variation of the hyperfine coupling constants~\cite{ZuchowskiPRL10,BruePRL12} or the interaction-induced second order spin-spin coupling~\cite{ZuchowskiCrYb}  (provided the widths of the Feshbach resonances are sufficiently broad). All investigated molecules can potentially be formed by using photoassociation but the actual pathways of the formation have to be investigated.  
To enhance molecule formation, STIRAP with atoms in a Mott insulator state produced by loading the BEC into an optical lattice~\cite{JakschPRL02} or nonresonant field control~\cite{GonzalezPRA12,TomzaMP13,TomzaPRL14} can be employed.

In the search for new kinds of ultracold molecules, the present paper draws attention to the highly magnetic and polar molecules formed from the highly magnetic lanthanide and $^1S$ or $^2S$ state atoms.
The results pave the way towards a more elaborate study of the magneto- or photoassociation and application of these polar and paramagnetic molecules at ultralow temperatures.
The present work also establishes the computational scheme for the future \textit{ab initio} investigations of the heteronuclear molecules containing the highly magnetic lanthanide atom with a large electronic orbital angular momentum such as erbium and dysprosium.

\acknowledgments
We would like to thank Tomasz Grining and Ludmi{\l}a Janion for many useful discussions.
Financial support 
from the Foundation for Polish Science within the START program is gratefully acknowledged.

\end{document}